\documentclass[onecolumn,11pt]{article}
\usepackage{epsfig}
\usepackage[usenames,dvipsnames]{pstricks}

\newcommand{\ket}[1]{|#1\rangle}
\title{Quantum mechanics\\ and the manifestation of the world%
\footnote{Paper presented at \emph{Berge Fest}, a conference celebrating the 60th birthday of Berge Englert (Centre for Quantum Technologies, National University of Singapore, 22--25 April 2014).}}
\author{Ulrich Mohrhoff\\
\textit{\small Sri Aurobindo International Centre of Education}\\ 
\textit{\small Pondicherry 605002 India}\\
\ttfamily{\small ujm@auromail.net}}
\date{}
\normalsize
\begin{document}
\maketitle
\begin{abstract}
\noindent Quantum theory's irreducible empirical core is a probability calculus. While it presupposes the events to which (and on the basis of which) it serves to assign probabilities, and therefore cannot account for their occurrence, it has to be consistent with it. It must make it possible to identify a system of observables that have measurement-independent values. What makes this possible is the incompleteness of the spatiotemporal differentiation of the physical world. This is shown by applying a novel interpretive principle to interfering alternatives involving distinctions between regions of space. Applying the same interpretive principle to alternatives involving distinctions between things makes it safe to claim that the macroworld comes into being through a progressive differentiation of a single, intrinsically undifferentiated entity. By entering into reflexive spatial relations, this entity gives rise to (i) what looks like a multiplicity of relata if the reflexive quality of the relations is not taken into account, and (ii) what looks like a substantial expanse if the spatial quality of the relations is reified. The necessary distinction between two domains (classical and quantum, or macro and micro) and their mutual dependence is best understood as a distinction between the manifested world and its manifestation.
\end{abstract}

\section{Introduction}
\label{intro}
There is no denying that quantum mechanics has a measurement problem. One would like to know why the theory's mathematical core is a probability calculus, and why the events it serves to correlate are measurement outcomes. But this is not what the usual interpretive fuss is about. What is commonly discussed as \emph{the} measurement problem is the problem of \emph{objectification}. The literature on this problem, which still follows largely its first rigorous formulations in the monographs of von Neumann~\cite{vonNeumann} and Pauli~\cite{Pauli1933}, postulates a measurement scheme comprised of three stages: the system preparation, a continuous dynamical process called ``premeasurement''  ($p$), and the seemingly miraculous appearance of an outcome called ``objectification'' ($o$):
\begin{equation}
\sum_{k}c_k\ket{A_0}\ket{q_k}\stackrel{(p)}{\longrightarrow}
\sum_{k}c_k\ket{A_k}\ket{q_k}\stackrel{(o)}{\longrightarrow}
\ket{A(q)}\ket{q}.
\label{trip}
\end{equation}
That there is something seriously wrong with this scheme can be deduced from the existence of proofs to the effect that the objectification problem is insoluble~\cite{Mittelstaedt,BLM}.

The usual interpretation of (\ref{trip}) is that initially the apparatus (really, objectively) is in the neutral state, and that eventually the apparatus (really, objectively) indicates that the measured observable has the value~$q$. To be sure, initially the apparatus is indeed in the neutral state, and eventually it indeed indicates a particular value, but this is not what these expressions mean.

What these expressions mean is that the initial state assigns probability 1 to the outcome of a measurement which indicates that the apparatus is in the neutral state, and the final state assigns probability 1 to the outcome of a measurement which indicates that the apparatus indicates the outcome~$q$. The quantum formalism itself does not warrant the assumption that probability~1 is sufficient for ``is'' or ``has.'' 

To make the prepared quantum state represent the \emph{fact} that the apparatus is in the neutral state, and to make the final quantum state represent the \emph{fact} that the apparatus indicates a particular outcome, one has to adopt the so-called ``eigenvalue-eigenstate link'' as an interpretive principle. Here is how this principle was formulated by Dirac~\cite{Dirac}:
\begin{quote}
The expression that an observable ``has a particular value'' for a particular state is permissible\dots in the special case when a measurement of the observable is certain to lead to the particular value, so that the state is an eigenstate of the observable.
\end{quote}
If we do not adopt the eigenvalue-eigenstate link, then we must distinguish between the collapse transition from $\sum_{k}c_k\ket{A_k}\ket{q_k}$ to $\ket{A(q)}\ket{q}$ and the objectification transition from probability~1 to ``is'' or ``has.'' In other words, even if we had an explanation for the collapse, it would not account for the fact that measurements tend to have outcomes. Nor could it, for there can be no dynamical explanation for an interpretive principle like the eigenvalue-eigenstate link.

\section{The real measurement problem}
\label{mmp}
Quantum mechanics presupposes the events to which (and on the basis of which) it assigns probabilities, and so it has to be consistent with their occurrence. But we cannot simply postulate that outcome-indicating devices are exempt from the quantum-mechanical correlation laws. We have to show that it is legitimate to look upon the values of certain observables as existing independently of measurements. 

Although a distinction has to be made between formulations and interpretations of quantum mechanics, the choice of a formulation cannot but bias the available interpretations. Among the better known formulations are Heisenberg's matrix formulation, Schr\"odinger's wave-function formulation, Feynman's path-integral formulation, the density-matrix formulation, and Wigner's phase-space formulation. I find it strange that a junior-level classical mechanics course devotes a considerable amount of time to different formulations of classical mechanics---such as Newtonian, Lagrangian, Hamiltonian, least action---whereas even graduate-level courses emphasize the wave-function formulation almost to the exclusion of all variants. Since this is the formulation that tempts us most to think of a quantum state as an evolving state that exists at every instant of time, and that implies, via the eigenvalue-eigenstate link, the actual existence of those properties or values to which probability~1 is assigned, we have every reason to distrust it.

To show that it is legitimate to look upon the values of certain observables as existing independently of measurements, I shall use Feynman's formulation, and I shall invoke an interpretive principle that seems natural in the context of this formulation, rather than interpretive principles like the eigenvalue-eigenstate link, which seem natural in the context of the wave function formulation.

Here goes. There are two core rules. The general task is to calculate the probability of a possible outcome of a final measurement, given the actual outcome of an initial measurement, and the way to proceed is to imagine a sequence of measurements that may be made in the meantime. If the intermediate measurements are actually made (or if it is possible to infer from other measurements what their outcomes would have been if they had been made), we use Rule~A, which requires us to first square the absolute values of the amplitudes and then add the results. If the intermediate measurements are not made (and if it is not possible to infer from other measurements what their outcomes would have been), we use Rule~B, which requires us to first add the amplitudes and then square the absolute value of the result.

From the wave-function point of view, Rule~B seems uncontroversial. Superpositions are ``normal,'' and what is normal requires no explanation. What calls for explanation is the existence of mixtures that admit of an ignorance interpretation. From Feynman's point of view, the uncontroversial rule is~A, inasmuch as this is what classical probability theory leads us to expect. What calls for explanation is why we have to add amplitudes, rather than probabilities, whenever the conditions stipulated by Rule~B are met. 

So why do we have to add amplitudes when we are required to do so? Here is why: 

\medskip\noindent\emph{Whenever quantum mechanics instructs us to use Rule~B, the distinctions we make between the alternatives correspond to nothing in the physical world. They lack objective reality.}

\medskip\noindent This is a statement about the structure or constitution of the physical world, not a statement merely of our practical or conceptual limitations.

\section{First application (regions of space)}
\label{firstapp}
Applied to a two-slit experiment with electrons (or any two-way interferometer for that matter), this interpretive principle implies that the distinction we make between ``the electron went through the left slit'' and ``the electron went through the right slit'' corresponds to nothing in the physical world. Since this distinction rests on a distinction between regions of space, it follows that space cannot be an intrinsically differentiated expanse. It has no ``inbuilt'' parts. But then what permits us to distinguish between ``here'' and ``there''? What furnishes space with its so-called parts?

What furnishes space with its so-called parts is its ``material content.'' A ``region of space'' can become the property of a particle (or atom, or molecule) only if it is realized by being monitored by a detector. (A detector is anything that can indicate the presence of something somewhere.) The apparatus is needed not only to indicate the possession of a property or a value, but also---and in the first place---to realize a set of properties or values so as to make them available for attribution. This is why the kinematical properties of microphysical objects only exist when their existence is indicated by macroscopic objects. To \emph{be} is to be \emph{measured}.

Now we are just a skip and a hop away from  identifying the observables whose values can legitimately be regarded as existing independently of measurements. Because an apparatus is needed to realize a region of space,  any attempt to partition space into progressively smaller regions will come to a halt when the detectors needed for the purpose cease to exist. We can therefore conceive of a partition of the physical world into \emph{finite} regions so small that none of them can be attributed (as a position) because none of them is realized (by a detector). In short, the spatial differentiation of the physical world is incomplete. It does not go ``all the way down.''

The same goes for the world's temporal differentiation, not only because of the relativistic interdependence of distances and durations but also because the indefiniteness principle for energy and time makes it impossible to realize sharp times. What is incomplete, therefore, is not quantum mechanics (as was argued by EPR) but the spatiotemporal differentiation of the objective world.

In an incompletely differentiated world, there will be objects whose position distributions are and remain so narrow that there are no detectors with narrower position distributions. If anything truly deserves the label ``macroscopic,'' it is these objects.

While arguments based on environmental decoherence cannot solve the objectification problem, or can solve it only for all practical purposes, they do support the conclusion that the indefiniteness of a macroscopic position is never revealed in the only way it could be revealed---through a departure from what the classical  laws predict. Macroscopic objects follow trajectories that are only counterfactually indefinite. Their positions are ``smeared out'' only relative to an imaginary spatiotemporal background that is more differentiated than the objective world. The testable correlations between the outcomes of measurements of macroscopic positions are therefore consistent with \emph{both} the classical \emph{and} the quantum laws. 

And so the observables we were looking for are the positions of macroscopic objects. We all knew this, of course. My point here was to show that it is consistent to do two things that are generally regarded as incompatible: to apply quantum mechanics to the positions of macroscopic objects, and to attribute to these positions a measurement-independent reality. What makes this possible is the incompleteness of the spatiotemporal differentiation of the physical world. (What renders the objectification problem insoluble is the complete temporal differentiation of the physical world implied by the tripartite measurement scheme inaugurated by von Neumann~\cite{vonNeumann} and Pauli~\cite{Pauli1933}.)

There are theorems in quantum field theory that extend the insolubility proofs for the objectification problem into the relativistic domain. Among them is a theorem by Clifton and Halvorson~\cite{CH}, to the effect that a particle cannot be ``in a state'' in which the probability of finding it in some finite region of space is~1. From this, Clifton and Halvorson have drawn the conclusion that the experience of detecting particles in finite regions of space is ``illusory'' and ``strictly fictional.''

What Clifton and Halvorson failed to take into account is that the spacetime manifold postulated by quantum field theory is not where experiments are performed. What is illusory is the notion that attributable positions are defined by spatial regions of this manifold. Attributable positions are defined by the sensitive regions of detectors, which, according to the aforesaid theorem, also cannot be localized in any finite region of space. What is strictly fictional is the spacetime manifold postulated by quantum field theory. What Clifton and Halvorson have shown is not that there are no localizable particles but that this manifold is not localizable relative to the positions that particles can possess.

\begin{figure}[t]
\begin{center}
\includegraphics[width=8.3cm]{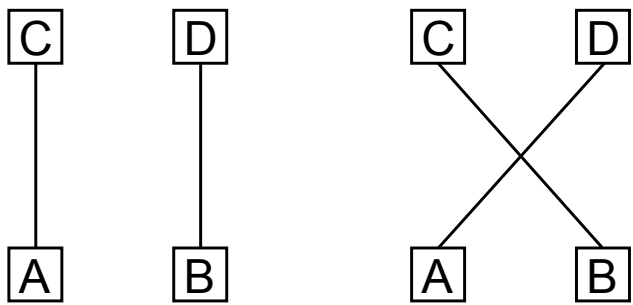}
\caption{Whenever quantum mechanics instructs us to add the amplitudes of the two alternatives rather than their probabilities, the straight lines, which represent transtemporal individuators of some kind, do not correspond to anything in the physical world.}
\label{fig-alt}
\end{center}
\end{figure}
\section{Second application (things)}
\label{secondapp}
The setup now consists of four non-overlapping regions $A$, $B$, $C$, $D$, realized by the sensitive regions of four detectors (Figure~1). Initial measurements indicate the presence of one particle in $A$ and one particle in~$B$. We wish to calculate the probability with which subsequently one particle is found in $C$ and one in~$D$. According to the interpretive principle introduced above, whenever quantum mechanics instructs us to add the corresponding amplitudes, the distinction we make between these two alternatives corresponds to nothing in the physical world. Since this distinction rests on a distinction between individual things, it follows that the particle found in $A$ is neither a different individual from the particle found in $D$ nor a different individual from the particle found in~$C$, and the same goes for the particle found in~$B$. One may therefore be excused for concluding that the particles in $C$ and $D$ (and hence the particles in $A$ and $B$ as well) are one and the same entity!

I am not the first to have this preposterous idea. In his Nobel Lecture  on December 11, 1965, Feynman recalled:
\begin{quote}
I received a telephone call one day at the graduate college at Princeton from Professor Wheeler, in which he said, ``Feynman, I know why all electrons have the same charge and the same mass.'' ``Why?'' ``Because, they are all the same electron!''
\end{quote}
What's more, there is no compelling reason to believe that this identity ceases to be real when it ceases to have observable consequences, owing to the presence of ``identity tags.'' Nothing therefore stands in the way of the claim that, intrinsically, each particle is \emph{numerically} identical with every other particle. What presents itself here and now with these properties and what presents itself there and then with those properties is the same entity.

What is certain, at any rate, is that the view according to which the world can be understood as a system of interacting constituents, has passed its expiration date. We cannot make sense of the behaviour of identical particles if we conceive of them as numerically distinct constituents. And if probability~1 is not sufficient for ``is'' or ``has,'' then the kinematical properties of microphysical objects only exist when they are indicated by---or can be inferred from---the behaviour of macroscopic objects, in which case the properties of macroscopic objects cannot be explained in terms of the properties of microscopic objects and their interactions. 

How are we to understand this mysterious interdependence of the two domains---macroscopic and microscopic, classical and quantum? Here is how it was stated by Landau and Lifshitz and by Michael Redhead:
\begin{quote}
quantum mechanics\dots contains classical mechanics as a limiting case, yet at the same time it requires this limiting case for its own formulation.~\cite{LL77}
\end{quote}
\begin{quote}
In a sense the reduction instead of descending linearly towards the elementary particles, moves in a circle, linking the reductive basis back to the higher levels.~\cite{Redhead1990}
\end{quote}
What Redhead appears to have missed, along with everyone else I should say, is that in addition to linking the properties of the quantum domain back to the properties of the classical domain, the alleged reduction does something else.

Here is what. Our attempts to conceptually divide the world reach a point where the distinctions we make between things, as well as the distinctions we make between regions of space, cease to exist. Instead of arriving at a ``reductive basis,'' we arrive at an intrinsically undivided spatial expanse and at an Entity that is numerically identical with every one of the world's so-called ultimate constituents. In other words, the real number of ultimate constituents is one. And since this single ultimate constituent is the only thing that exists independently of anything else, we would be justified in calling it Pure Being, or something to that effect.

\section{Manifestation}
\label{manifestation}
The reason why it is so hard to beat sense into quantum mechanics is that the theory answers a question that we are not in the habit of asking. The question we should be asking is this: How are \emph{forms}---the shapes of things---\emph{manifested}? This question has a simple answer: forms (including the all-encompassing form called ``space'') are manifested by means of \emph{reflexive spatial relations}. By entering into reflexive spatial relations, Pure Being gives rise to (i)~what looks like a multiplicity of relata if the reflexive quality of the relations is ignored, and (ii)~what looks like a substantial expanse if the spatial	quality of the relations is reified. As Leibniz said, \emph{omnibus ex nihilo ducendis sufficit unum}---one is enough to create everything from nothing. A single self-existent entity is enough to create the relata we call particles as well as the expanse we call space.

The distinction between the two domains is essentially the distinction between the manifested world and its manifestation. While the manifested world is a world of interacting objects and causally connected events, the manifestation of such a world cannot be understood in terms of what is manifested. Instead of describing it in the language of interacting objects and causally connected events, quantum mechanics describes it in the language of \emph{correlations} between the possible outcomes of hypothetically performed measurements. This is how the alleged reduction ``moves in a circle, linking the reductive basis back to the higher levels.''

The manifestation of the macroworld is a transition from a condition of complete indefiniteness and indistinguishability to a condition of complete definiteness and distinguishability, and what is not definite or distinguishable can only be described in terms of correlations between events that are definite and distinguishable. What is instrumental in the manifestation can only be described in terms of what is manifested.

If what we have is a Pure Being capable of entering into reflexive spatial relations, and if what we want is a world of deterministically evolving objects, then we need spatially extended objects that are reasonably stable. How does one create such objects using nothing but spatial relations between relata that lack spatial extent? The answer to this question is: quantum mechanics. 

In a previous paper~\cite{Mohrhoff-QMexplained} I derived quantum mechanics from the existence of reasonably stable, spatially extended objects that are (or appear to be) composites of objects that lack spatial extent. What is crucial for the existence of such objects is the indefiniteness of their internal relative positions and momenta. This raises the question of what is the right or best way of dealing with an indefinite physical quantity. And the answer to this question is: assigning probabilities to the possible outcomes of a measurement of this quantity. This is one reason why the mathematical formalism of quantum physics is a probability calculus, and why the events to which it serves to assign probabilities are measurement outcomes.

It may even be that quantum mechanics, in turn, requires the validity of both the standard model and general relativity. For (i)~quantum mechanics presupposes the macroscopic events which it correlates, (ii)~the existence of macroscopic events requires a sufficient variety of chemical elements, and (iii)~the existence of a sufficient variety of chemical elements requires the validity of the standard model and general relativity, at least as effective theories. In another paper~\cite{Mohrhoff-justso} I put forward arguments in support of these claims. (See also Chap.\ 22 of Mohrhoff~\cite{Mohrhoff-book}.)

\section{Summary}
\label{summary}
What quantum mechanics is trying to tell us about the physical world is that there is an Ultimate Reality---undifferentiated and therefore beyond categorization---which manifests the world by entering into reflexive spatial relations. The relations constitute space, and the resulting apparent multitude of relata is what, for want of a better word, we call particles.

If a ``classical'' world emerges, it is not from some mystical domain of potentiality, nor by a dynamical process, nor through environmental decoherence, but by a progressive manifestation. The first stage of this atemporal process, which is probed by high-energy physics, can only be described in terms of correlations between the ``clicks'' of non-existent detectors. At energies low enough for atoms to be stable, it becomes possible to think in terms of objects with fixed numbers of components. These we describe in terms of correlations between the possible outcomes of unperformed measurements. 

Molecules bring us another step closer to the manifested world, inasmuch as they are the first objects with forms that can be visualized---their atomic configurations. But it is only the finished product---the macroworld---that gives us the actual detector ``clicks'' and the actual measurement outcomes that allow us to test the correlations in terms of which quantum mechanics describes the manifestation of the world---that is, the transition from Pure Being to the manifested world.

Physical theory proceeds from what is obvious. What is obvious to anyone who can tell the difference between mathematics and the physical world is that physics concerns correlations between measurement outcomes. What is also obvious by now is the atomic constitution of matter, the stability of matter under conditions favourable to the existence of physicists, and the dependence of the stability of matter on the stability of atoms. From these data one can derive quantum mechanics~\cite{Mohrhoff-QMexplained} and, arguably, the standard model and general relativity,  at least as effective theories~\cite{Mohrhoff-justso}. The bottom line: Fundamentalism in physics is a red herring. While we are (or may be) able to deduce QM+SM+GR from what is obvious, we are not in a position to deduce what is obvious from any physical theory.

Before concluding I need to address ``some technical quibbles'' raised by a reviewer. Noting that I seem to assume that detectors are necessarily macroscopic, the reviewer requests that this be stated as an assumption, and further asks whether the incompleteness of the spatial differentiation of the physical world implies the existence of a ``fundamental length.'' The question whether detectors are necessarily macroscopic obviously depends on the respective definitions of ``detector'' and ``macroscopic.'' A detector, the reader will recall, is anything that can indicate the presence of something somewhere, and a macroscopic object is an object $M$whose position distribution (determined by all outcome-indicating events that contribute to define it) is and remains so narrow that there are no detectors with narrower position distributions---detectors that could distinguish between regions over which $M$'s position is distributed.

The context in which this question arises is the challenge to demonstrate what von Weizs\"acker has called ``semantic consistency'': ``Semantic consistency of a theory will mean that its preconceptions, how we interpret the mathematical structure physically, will themselves obey the laws of the theory''~\cite{vW}.  To demonstrate the semantic consistency of quantum mechanics means to establish the consistency of the theory's formal apparatus (qua probability calculus) with the outcome-indicating events whose existence it presupposes and which it serves to correlate---in short, to establish the consistency of the correlations with their correlata. That this is problematic is shown by the insolubility proofs for the objectification problem and their extensions into the relativistic domain. 

What makes this problematic is the all but universally accepted view that the points and instants on which a wave function depends correspond one-to-one to the elements of an intrinsically and completely differentiated spacetime. Establishing the semantic consistency of quantum mechanics is made possible by the interpretive principle introduced in Sect.~\ref{mmp}, inasmuch as this implies, via the incompleteness of the spatiotemporal differentiation of the physical world, the existence of macroscopic objects, which follow trajectories that are only counterfactually indefinite, so that the testable correlations between the outcomes of measurements of macroscopic positions are consistent with \emph{both} the classical \emph{and} the quantum laws. Because these (and only these) correlations are consistent with the classical laws, we can attribute to those (and only those) positions a measurement-independent reality, and this not merely for all practical purposes. Non-macroscopic positions only exist (or have values) if and when (and to the extent that) they are measured. Only a macroscopic position, therefore, is capable of indicating a measurement outcome. And since a detector is something that can indicate the presence of something somewhere, it must possess a macroscopic part---the proverbial pointer needle---whose position can serve to indicate the detection of an object in the detector's sensitive region. That detectors are necessarily macroscopic (in this sense) is therefore not an independent assumption but a necessary consequence of the interpretive principle introduced in Sect.~\ref{mmp}.

As to the question whether the incompleteness of the spatial differentiation of the physical world implies the existence of a fundamental length, I don't see how this might be the case.

It goes without saying that the novel interpretation of quantum mechanics presented here raises further questions and/or objections. Some of these are answered in another paper~\cite{Mohrhoff-manifesting}---owing to its length somewhat to the detriment of the overall picture, which the present more succinct paper aims to highlight.

\end{document}